\documentstyle[editedvolume]{crckapb}

\newcommand{\apj}{ApJ}
\newcommand{\aj}{AJ}

\newcommand{\mvrr}{\hbox {$\rm M_v(RR)$}}
\newcommand{\ea}{{\it et al.}}
\newcommand{\feh}{\hbox{$ [{\rm Fe}/{\rm H}]$}}
\newcommand{\laa}
{\mathrel{\hbox{\rlap{\hbox{\lower4pt\hbox{$\sim$}}}\hbox{$<$}}}}
\newcommand{\gaa}
{\mathrel{\hbox{\rlap{\hbox{\lower4pt\hbox{$\sim$}}}\hbox{$>$}}}}

\begin{opening}
\title{GLOBULAR CLUSTER DISTANCE DETERMINATIONS}

\author{BRIAN CHABOYER}
\institute{Steward Observatory, University of Arizona\\ 
Tucson, AZ, USA and\\
Department of Physics and Astronomy, Dartmouth College\\ Hanover, NH, USA}

\end{opening}

\runningtitle{GLOBULAR CLUSTER DISTANCES}

\begin{document}

\section{Introduction}
The distance scale to globular clusters is of great interest for two
principal reasons.  Firstly, the best estimates for the 
absolute ages of globular clusters require that the distance to the
clusters be known (e.g.\ Renzini 1991; Chaboyer 1996).  These age
determinations provide the best estimate for the age of the universe,
but are very sensitive to the adopted globular cluster distance scale.
For example, a revision in the distance scale by 0.10 mag changes the
derived ages by 10\%.  The second reason for the interest in the
globular cluster distance scale is that globular clusters contain
stellar populations (RR Lyrae stars, tip of the red giant branch)
which are commonly used as distance indicators in astronomy.  Thus,
globular clusters can serve as nearby calibrators of these standard
candles.

The release of the Hipparcos data set has caused a number of workers
to re-examine the question of the globular cluster distance scale.
Hipparcos provided high quality parallaxes for a number of nearby
metal-poor stars, yielding a calibration of the absolute magnitude of
metal-poor stars which could be used to derive the distance to
globular clusters from main sequence fitting (Reid 1997, 1998;
Gratton \ea\ 1997; Chaboyer \ea\ 1998; Grundahl \ea\ 1998; Pont \ea\
1998).  In addition, Hipparcos provided proper motions for a large
number of RR Lyrae stars.  These proper motions have been combined
with the method of statistical parallax to
estimate the absolute magnitude of RR Lyrae stars, one of the standard
candles found in globular clusters (Fernley \ea\ 1998a). 

In this chapter, the results from these studies and other recent
investigations of the globular cluster distance scale will be
reviewed.  This is not meant to be a comprehensive review, as only
results from the last few years are discussed in detail.  
The calibration of the absolute magnitude of the RR Lyrae
stars is presented in \S \ref{sect1}.  Astrometric distances derived 
from internal proper motion and  radial velocity studies are discussed in
\S \ref{sect2}.  Section \ref{sect3} contains a summary of results
based upon main sequence fitting; a more complete discussion may be
found in the chapter by Gratton \ea\ in this volume.   The results of
white dwarf sequence fitting are presented in \S \ref{sect4}, while
the potential of other distance indicators is discussed in \S \ref{sect5}.
The various results are compared in \S \ref{sect7} which summarizes
the current status of the globular cluster distance scale.

\section{RR Lyrae stars \label{sect1}}
RR Lyrae stars are radially pulsating variable stars, which have
traditionally been used as standard candles in astronomy. As RR Lyrae
stars are found in many globular clusters, they are one of the primary
distance indicators to globular clusters.  However, RR Lyrae stars are
not perfect standard candles; it has been known for many years that
their absolute magnitude (\mvrr) is a function of metallicity (Sandage
1981a,b). This has traditionally been parameterized as a simple linear
relationship between \mvrr\ and \feh:
\begin{eqnarray}
\mvrr\ = \alpha\,\feh  +  \beta.
\label{eqmvrr}
\end{eqnarray}
There are a variety of different techniques which can be used to
determine \mvrr , and hence calibrate RR Lyrae stars as standard
candles to be used in determining the distances to globular clusters.
Some of these methods are best for ascertaining the  zero-point of the
\mvrr-metallicity relation ($\beta$), while others are best in
determining the variation of \mvrr\ with metallicity ($\alpha$).  The
determination of each of these quantities will be discussed in turn.

\subsection{Variation of \mvrr\ with metallicity }
Sandage (1981a,b) first derived the coefficients in equation
(\ref{eqmvrr}) empirically, and estimated a ``steep" slope $\alpha$ of
0.35.  From a theoretical calibration based on synthetic horizontal
branch (HB) population models, Lee \ea\ (1990, 1994) derived a ``shallower"
$\alpha$ in the range 0.17-0.19.  It is important to note that
theoretical models predict that a simple linear relationship between
\mvrr\ and \feh\ does not exist. Theory predicts that stars of the
same metallicity evolve through the RR Lyrae instability strip at
different luminosities depending on whether they originate on the red
or blue side of the instability strip.  Thus \mvrr\ depends on
HB morphology.  At a given metallicity, RR Lyrae
variables are more luminous in clusters with blue HB
morphology types than in red HB morphology type (Lee
1991).  This difficulty suggests that the standard \mvrr\ calibration
should not be applied to globular clusters with extremely blue
HB morphologies (e.g.\ see Fig.\ 1 in Caputo 1997).
Furthermore, theoretical HB models do not predict a
simple linear relationship between \mvrr\ and \feh\ even among
globular clusters with similar HB types (Caputo 1997;
Caloi \ea\ 1997).  For example, with $[\alpha/{\rm Fe}] = +0.4$,
Caputo (1997) predicts $\alpha=0.19$ for $\feh < -1.6$ and $\alpha =
0.32$ for $\feh > -1.6$.  The large differences between these two
slopes, suggest that the traditional assumption of a simple linear
relationship between \mvrr\ and \feh\ is not valid.  This is certainly
true in detail, but a global fit over the range of metallicities
typically found in globular clusters with RR Lyrae stars ($-2.2 \le
\feh \le -1.0$) finds $\alpha = 0.25$, with a maximum deviation of
0.02 mag in \mvrr.  Even a fit over a very broad metallicity range 
($-2.2 \le \feh \le 0.0$) leads to maximum deviations of only 0.04 mag
in \mvrr\ between a simple linear fit and the relationship derived
directly from the models.  Given the small residuals between the linear fit
and the \mvrr\ values predicted by theoretical models, it is justified
to assume a linear relation between \mvrr\ and \feh\ for distance
determinations.  

The semi-empirical Baade-Wesselink method supports a shallow slope,
$\alpha = 0.20\pm 0.04$ (Fernley \ea\ 1998b).  Note that the
Baade-Wesselink results include many high metallicity points
($\feh \ge 0.5$) and the theoretical models would predict a bias to a
higher slope.  The shallow slope is also supported by HST observations
of globular cluster HBs in M31, where $\alpha =
0.13\pm 0.07$ (Fusi Pecci \ea 1996).  The clusters span a range in
metallicity ($-1.8 \le \feh \le -0.4$) which is unlikely to introduce
a significant bias to the derived value of $\alpha$.  Using the
relation between Fourier decomposition and luminosity for RRab stars
in globular clusters Kov\'{a}cs \& Jurcsik (1996) determined that
$\alpha$ is less than 0.20.  Overall, the observations and theoretical
models appear to favor somewhat shallow slopes ($\alpha \le 0.26$).
In the rest of this review a value of $\alpha = 0.23\pm 0.04$ is
adopted.  The 1-$\sigma$ range of this value (0.19 --- 0.27)
encompasses the majority of recent determinations of $\alpha$.

\subsection{Zero-point of \mvrr\ with metallicity}
\subsubsection{Statistical Parallax}
A traditional method used to determine the absolute magnitude of RR
Lyrae stars is statistical parallax (see Layden this volume). This
method determines the absolute magnitude of RR Lyrae stars in the
field.  Hipparcos obtained a large number of proper motions which can
be used in the statistical parallax solution.  Current statistical
parallax solutions find $\mvrr = 0.77\pm 0.13$ mag at $<\feh > = -1.60$
(Gould \& Popowski 1998; also Layden, this volume).  Combining this
with the estimate for the slope given in the previous section yields 
\begin{eqnarray}
\mvrr\ = (0.23\pm 0.04)(\feh + 1.6) +  (0.77\pm 0.13).
\label{statpi}
\end{eqnarray}

\subsubsection{Calibration via the LMC}
Given a distance estimate to the LMC, the observed magnitude of RR
Lyrae stars in the LMC can be used to calibrate \mvrr\
(e.g.\ Walker 1992).  Walker (this volume) summarizes current distance
estimates to the LMC, and concludes that the distance modulus to the
LMC is $18.55\pm 0.10$ mag.  Walker (1992) determined the mean
magnitude of a large number of RR Lyrae stars in several clusters in
the LMC.  Combining this data with the above distance modulus to the
LMC yields $\mvrr = 0.39\pm 0.10$ mag at $<\feh> = -1.90$. With $\alpha =
0.23\pm 0.04$ this yields
\begin{eqnarray}
\mvrr\ = (0.23\pm 0.04)(\feh + 1.6) +  (0.46\pm 0.11)
\label{lmc}
\end{eqnarray}
(allowing for an error of 0.15 dex in the mean LMC \feh).  A
comparison between equations (\ref{statpi}) and (\ref{lmc}) indicates
that these two methods for determining the zero-point of the
\mvrr-\feh\ relation differ by $1.8\,\sigma$.

\subsubsection{Theoretical HB Models}
Theoretical stellar evolution models may be used to derive the
absolute magnitude of the zero-age horizontal branch (ZAHB).  It is
important to note that the results of these calculations depend
sensitively on the assumed helium abundance used in the calculations,
along with the physics used in the construction of the stellar models.
A change in the assumed main sequence helium abundance by 4\% (from
$Y=0.23$ to $Y=0.24$ for example) leads to a change in the predicted
HB luminosity of approximately 0.05 mag. Cassisi \ea\ (1998) show that
improvements in the physics used in the theoretical models over the
last 10 years has lead to an increase in the predicted ZAHB luminosity
by about 0.15 mag.  For this reason, only globular cluster distance
determinations based upon the latest input physics will be considered
in this subsection.

A number of authors have used theoretical ZAHB models to derive the
distance to specific globular clusters.  Brocato \ea\ (1997)
constructed ZAHB models for M68, and compared these to the observations
obtained by Walker (1994). The existence of a blue tail on the M68 HB
allowed Brocato \ea\ (1997) to derive the distance and reddening to
M68 simultaneously.  They obtained ${\rm (m - M)_V}= 15.25$ mag and ${\rm
E(B-V)} = 0.05$ using their most recent models.  Walker (1994)
obtained a mean apparent magnitude of $V=15.67\pm 0.04$ mag for the RR
Lyrae stars in M68.  Estimates of the metallicity of M68 vary from
$\feh = -2.17$ (Minniti \ea\ 1993) to $\feh = -1.99$ (Carretta \&
Gratton 1997). Taking the average of these two metallicity 
estimates, the distance
modulus derived by Brocato \ea\ (1997) implies $\mvrr = 0.42$ mag at $\feh =
-2.08$.  

Salaris \ea\ (1997) performed a fit to M68 using
their ZAHB models and isochrones.  The distance modulus
and reddening were determined by shifting the ZAHB models and
isochrones in order to match the observed main-sequence ridge line and
the ZAHB level in the RR Lyrae region.  Salaris \ea\ (1997)
obtained ${\rm (m -M)_V}= 15.26$ mag and ${\rm E(B-V)} = 0.06$.  This
implies $\mvrr = 0.41$ mag, which is very similar to the result obtained
by Brocato \ea\ (1997).  

Finally, Caloi \ea\ (1997) have used their ZAHB models to determine
the distance to three globular clusters.  Their models differ from
other workers in that they do not use mixing length theory, but 
adopt the Canuto \& Mazzitelli (1991) convection treatment.  Using the same 
M68 data, Caloi \ea\ (1997) determined a distance modulus of 
${\rm (m - M)_V}=15.37$ mag (assuming ${\rm A_V = 3.2\,E(B-V)}$).  This
is  0.11 mag larger than the value found by Brocato \ea\ (1997) and Salaris
\ea\ (1997).  However, the work of Caloi \ea\ (1997) ignored the
fact that the $\alpha$ capture elements are enhanced over their solar
ratio in metal poor stars (e.g.\ Nissen \ea\ 1994).  Caloi \ea\ (1997)
used their $Z=0.0001$ models to compare to M68, while Salaris \ea\
(1997) and Brocato \ea\ (1997) take into account $\alpha$ element
enhancement by using their $Z=0.0002$ models. The results of Caloi
\ea\ (1997) may be corrected to include $\alpha$ element enhancement
with the aid of their Table 2.  Performing such a correction leads to
${\rm (m - M)_V}=15.28$ mag, implying $\mvrr = 0.39$ mag in good
agreement with Brocato \ea\ (1997) and Salaris \ea\ (1997). Averaging
these three determinations for the distance to M68 yields $\mvrr =
0.41$ mag at $\feh = -2.08$ based upon theoretical ZAHB models.

Caloi \ea\ (1997) also derived the distance to M5.  Once again,
correcting their published value for the effects of $\alpha$ element
enhancement leads to a distance modulus of ${\rm (m - M)_V} = 14.51$ mag
for M5.  M5 has a metallicity of $\feh = -1.17$ from high dispersion
spectroscopic analysis (Sneden \ea 1992) and mean RR Lyrae apparent
magnitude of $V=15.05\pm 0.06$ mag (Reid 1996). Thus, the theoretical ZAHB
models of Caloi \ea\ (1997) imply $\mvrr = 0.56$ mag at $\feh = -1.17$.
This may be combined with the M68  calibration above to yield a
calibration of \mvrr\ based upon theoretical ZAHB models from three
different groups:
\begin{eqnarray}
\mvrr\ = (0.23\pm 0.04)(\feh + 1.6) +  (0.49\pm 0.10).
\label{hbtheory}
\end{eqnarray}
The error in the zero-point has been estimated from a
consideration of the uncertainties associated with the theoretical HB
models, discussed in the beginning of this section.

\section{Astrometric Distances \label{sect2}}
A comparison of the proper motion and radial velocity dispersions
within a cluster allows for a direct determination of GC distances,
independent of reddening (Cudworth 1979).  Although this method
requires that a dynamical model of a cluster be constructed, it is the
only method considered here which directly measures the distance to a
GC without the use of a `standard' candle.  The chief disadvantage of
this technique is its relatively low precision.  This problem is
avoided by averaging together the astrometric distances to a number of
different GCs.  Rees (1996) presents new astrometric distances to
eight GCs, along with two previous determinations.  As pointed out by
Rees, there are possibly large systematic errors in the dynamical
modeling of M15, NGC 6397 and 47 Tuc. As such, these clusters will be
excluded from our analysis.  Rees (private communication) has performed
a new reduction of the M2 proper motions, yielding
a total of seven clusters whose distances have been estimated
astrometrically.  

Table \ref{tabastro} tabulates the astrometric distances from Rees
(1996) along with the new distance determination to M2.  Unless
otherwise noted, the numbers are those given by Rees (1996).  For the
\feh\ values, preference has been given to the high dispersion results
of Kraft, Sneden and collaborators.  Table
\ref{tabastro} also includes the HB type of the clusters
taken from Harris (1996).  This is defined to be $(B-R)/(B+V+R)$,
where $B$, $V$ and $R$ are the numbers of blue, variable and red HB
stars.  Taking the weighted average of the \mvrr\ values listed in
Table \ref{tabastro} results in $\mvrr = 0.60\pm 0.10$ mag at $<\feh> =
-1.60$, where the average \feh\ value has been calculated using the
same weights as in the \mvrr\ average.  This implies
\begin{eqnarray}
\mvrr\ = (0.23\pm 0.04)(\feh + 1.6) +  (0.60\pm 0.10)
\label{astrometric}
\end{eqnarray}
using the \mvrr-\feh\ slope adopted in \S 2.1.
\begin{table}[htb]
\begin{center}
\begin{minipage}{11.06cm}
\caption{Astrometric Distances}
\label{tabastro}
\begin{tabular}{llccll}
\hline
\multicolumn{1}{c}{Cluster}&
\multicolumn{1}{c}{\feh}&
\multicolumn{1}{c}{HB Type}&
\multicolumn{1}{c}{${\rm (m - M)_O}$}&
\multicolumn{1}{c}{V(HB)} & 
\multicolumn{1}{c}{$\rm M_V(HB)$}  \\
\hline
M5\footnote{\feh\ from Sneden \ea\ (1992).}
   &   $-1.17$ & $+0.31$ & 14.44 & $15.05$  &   $0.51 \pm  0.41$\\
M4\footnote{\feh\ from Zinn \& West (1984).}
   &   $-1.33$ & $-0.06$ & 11.18 & $13.37$  &   $0.67 \pm  0.23$\\
M3\footnote{V(HB) from Buonanno \ea\ (1994). Reddening
from Zinn (1985). \feh\ \hspace*{1cm} from Kraft \ea\ (1992).} 
   &   $-1.47$ & $+0.08$ & 14.91 & $15.63$  &   $0.69 \pm  0.59$\\
M13\footnote{V(HB) from Buonanno \ea\ (1989).  \feh\ 
from Kraft \ea\ (1997)}
  &   $-1.58$ & $+0.97$ & 14.06 & $14.83$  &   $0.71 \pm  0.23$\\
M2\footnote{V(HB) from Harris (1996).  \feh\ from Zinn \& West (1984)}
   &   $-1.62$ & $+0.96$ & 15.26 & $16.05$  &   $0.63 \pm  0.25$\\
M22$^b$  &   $-1.75$ & $+0.91$ & 12.17 & $14.10$  &   $0.58 \pm  0.19$\\
M92\footnote{\feh\ from Sneden \ea\ (1991).}
  &   $-2.25$ & $+0.91$ & 14.76 & $15.13$  &   $0.31 \pm  0.32$\\
\hline
\end{tabular}
\end{minipage}
\end{center}
\end{table}

The four most metal-poor clusters all have very blue HB types.  For
these clusters, theoretical models suggest that the RR Lyrae stars
will be more luminous than for stars with redder HB types.  An average
of the blue HB clusters finds $\mvrr = 0.59\pm 0.12$ at $<\feh> =
-1.71$, while the three clusters with redder HB types yield $\mvrr =
0.64\pm 0.19$ at $<\feh> = -1.31$.  Translating these two estimates to
$\feh = -1.60$ (using $\alpha = 0.23\pm 0.04$) yields $\mvrr = 0.62\pm
0.12$ for the blue HB clusters and $\mvrr = 0.57\pm 0.19$ for the
other clusters.  There does not appear to be a significant difference
between the two \mvrr\ calibrations, and so the averaging used to
derive equation (\ref{astrometric}) appears to be valid.

\section{Main Sequence Fitting \label{sect3}}
Hipparcos provided high quality parallaxes for a number of metal-poor
field stars.  This has prompted a number of authors to determine new
distances to globular clusters using main sequence fitting.  Main
sequence fitting is discussed in detail in the chapter by Gratton \ea\
in this book.  The results of the published investigations are
summarized in Table \ref{mainfit}.  The typical distance modulus
errors quoted by the various authors is $\pm 0.10$ mag.  The authors
took quite different approaches in dealing with issues such as sample
selection, reddening, biases, etc.  In general, the distance moduli
derived by various authors for a given globular cluster are in good
agreement. For example, the various distance modulus estimates to M13
agree to within $\pm 0.03$ mag. The Grundahl \ea\ (1998) distance
estimate to M13 is particularly noteworthy as they utilized
Str\"{o}mgren photometry, while the other authors used B,V photometry.
Of course, all of these work utilize the same basic assumption, that
the nearby metal-poor stars have identical properties to their
metal-poor counterparts in globular clusters.  
\begin{table}[htb]
\begin{center}
\begin{minipage}{11.55cm}
\caption{Main Sequence Fitting Distances}
\label{mainfit}
\begin{tabular}{llllc}
\hline
\multicolumn{1}{c}{Cluster}&
\multicolumn{1}{c}{${\rm (m-M)_O}$}&
\multicolumn{1}{c}{${\rm E(B-V)}$}&
\multicolumn{1}{c}{${\rm (m - M)_V}$}&
\multicolumn{1}{c}{Reference}\\ 
\hline
47 Tuc NGC 104 &13.56  &  0.04   &     13.69  &   1\\[-2pt]
               &13.44  &  0.055  &     13.62  &   2\\[5pt]
 	                   	            
NGC 288        &15.00  &  0.01   &     15.03  &   1\\[-2pt]
               &14.83  &  0.033  &     14.94  &   2\\[5pt]
	                   	            
NGC 362        &14.86  &  0.056  &     15.04  &   2\\[5pt]
	                   	            
M68 NGC 4590   &15.18  &  0.040  &     15.31  &   2\\[5pt]
	                   	            
M5 NGC 5904    &14.52  &  0.02   &     14.58  &   1\\[-2pt]
               &14.41  &  0.03   &     14.51  &   3\\[-2pt]
               &14.49  &  0.035  &     14.60  &   2\\[5pt]
	                   	            
M13 NGC 6205   &14.38  &  0.021  &     14.45  &   4\\[-2pt]
               &14.45  &  0.02   &     14.51  &   1\\[-2pt]
               &14.41  &  0.02   &     14.47  &   3\\[-2pt]
               &14.39  &  0.020  &     14.45  &   2\\[5pt]
	                   	            
M92 NGC 6341   &14.72  &  0.025  &     14.80  &   2\\[-2pt]
               &14.68  &  0.02   &     14.74  &   5\footnote{This is
the distance modulus derived by Pont \ea\ when they do not include the
known binaries in their fit.}\\[5pt]
	                   	            
NGC 6397       &12.24  &  0.19   &     12.85  &   1\\
	                   	            
NGC 6752       &13.16  &  0.04   &     13.29  &   1\\[-2pt]
               &13.20  &  0.04   &     13.33  &   3\\[-2pt]
               &13.21  &  0.035  &     13.32  &   2\\[5pt]
	                   	            
M71 NGC 6838   &13.19  &  0.28   &     14.09  &   1\\[5pt]
	                   	            
M30 NGC 7099   &14.82  &  0.039  &     14.94  &   2\\

\hline
\end{tabular}

{\small REFERENCES. --- (1) Reid 1998; (2) Gratton \ea\ (1997); (3)
Chaboyer \ea\ (1998); (4) Grundahl \ea\ (1998); (5) Pont \ea\ (1998).}

\end{minipage}
\end{center}
\end{table}

Some of the globular clusters listed in Table \ref{mainfit} have very
good RR Lyrae mean magnitudes, and so the main sequence fitting
distances may be compared amongst each other, and to other methods
using \mvrr\ (equation \ref{eqmvrr}).  For example M92 has a mean RR
Lyrae magnitude of $V=15.10\pm 0.03$ mag (Carney \ea\ 1992).  Averaging
the distance moduli obtained by Gratton \ea\ (1997) and Pont \ea\
(1998) yields $\mvrr = 0.33\pm 0.10$ mag (at $\feh = -2.25$ from Sneden
\ea\ 1991).   The Gratton \ea\ (1997) distance modulus to M68  yields 
$\mvrr = 0.36\pm 0.10$ mag (at $\feh = -2.08$ using the data for M68 given
in \S 2.2.3).  These two estimates for \mvrr\ can be directly compared
at an intermediate metallicity ($\feh = -2.16$) using equation
\ref{eqmvrr} which yields $\mvrr = 0.35\pm 0.10$ mag for M92 and 
$\mvrr = 0.34\pm 0.10$ mag for M68.  Note that M92 has a blue HB (HB-type
of +0.91), while the M68 has a much redder HB (HB-type of 0.17).  This
comparison indicates,  that for these two clusters the HB type does
not have a significant effect on \mvrr.

Good RR Lyrae photometry also exists for M5 (see references in \S
2.2.3).  Averaging together the three main sequence fitting results
for M5  presented in Table \ref{mainfit} results in ${\rm (m-M)_V} = 14.56\pm
0.10$ mag and $\mvrr = 0.49\pm 0.10$ mag at $\feh = -1.17$.  Taking the mean 
determination of \mvrr\ for M92, M68 and M5 from main
sequence fitting (and using $\alpha = 0.23\pm 0.04$ in equation
\ref{eqmvrr}) yields
\begin{eqnarray}
\mvrr\ = (0.23\pm 0.04)(\feh + 1.6) +  (0.45\pm 0.10).
\label{eqmainfit}
\end{eqnarray}

\section{White Dwarf Fitting \label{sect4}}
Renzini \ea\ (1996) have utilized deep HST WFPC2 observations of NGC
6752 to obtain accurate photometry of the cluster white dwarfs.  In
addition, they obtained similar photometry of nearby white dwarfs
which appear to have similar masses to the cluster white dwarfs.
Using the parallaxes of the nearby white dwarfs to determine their
absolute magnitude, they determined the distance to NGC 6752 using a
procedure similar to main sequence fitting.  The key assumption in
this method is that the masses of the local white dwarfs are similar
to the masses of the white dwarfs in NGC 6752.  The derived distance
modulus is ${\rm (m - M)_V} = 13.18\pm 0.10$ mag assuming ${\rm
E(B-V)} = 0.04$.  This reddening estimate is from Zinn (1985), and is
identical to those found by Burnstein \& Heiles (1982) and Carney
(1979).

The average distance modulus for NGC 6752 from main sequence fitting
is ${\rm (m - M)_V} =13.31\pm 0.10$ mag (Table \ref{mainfit}), leading
to a difference of 0.13 mag between the main sequence and white dwarf
fitting distance estimates to NGC 6752.  This cluster has a very blue
HB, and so determination of its HB magnitude at the position of the RR
Lyrae instability strip is very difficult.  In order to compare white
dwarf fitting to the other distance determination techniques, equation
(\ref{eqmainfit}) can be combined with the difference between the white
dwarf and main sequence fitting distances to NGC 6752 to yield
\begin{eqnarray}
\mvrr\ = (0.23\pm 0.04)(\feh + 1.6) +  (0.58\pm 0.10).
\label{eqwhitefit}
\end{eqnarray}

\section{Other Distance Indicators \label{sect5}}
There are a variety of other methods which have been used to obtain
distances to globular clusters.  Jimenez \& Padoan (1998) have
compared theoretical luminosity functions to observed luminosity
functions of M5 and M55.  For M5, they determined 
${\rm (m-M)_V} = 14.55\pm 0.10$ mag. This can be compared to the average
distance modulus derived from main sequence fitting  of 
${\rm (m-M)_V} = 14.56\pm 0.10$ mag (Table \ref{mainfit}).  

Kov\'{a}cs and Walker (1998) have presented a detailed analysis of
double-mode RR Lyrae stars in M15, M68 and IC 4499.  This analysis is
based upon linear pulsation models and is free of systematic effects
due to ambiguities in the various zero-points (bolometric corrections,
magnitudes, etc).  The derived absolute magnitudes are 0.2 -- 0.3 mag
brighter than corresponding Baade-Wesselink values which are tied to
the statistical parallax zero-point.  

Simon \& Clement (1993) used hydrodynamic pulsation models to show
that physical properties (such as absolute magnitude) of RRc stars
could be derived from their pulsation period and Fourier phase
parameters.  Kaluzny \ea\ (1998) present \mvrr\ for seven globular
clusters based upon this method.  For example, for M68, they find 
$\mvrr = 0.38$ mag, which compares to $\mvrr = 0.41$ mag from theoretical HB
models (\S 2.2.3) and $\mvrr = 0.36$ mag from main sequence fitting (\S
\ref{sect4}).  For M5, Kaluzny \ea\ (1998) tabulate $\mvrr = 0.61$ mag
which agrees well with the theoretical HB models ($\mvrr = 0.56$ mag) and
is  somewhat fainter than that derived from main sequence fitting
($\mvrr = 0.49$ mag).  

The discovery of a detached eclipsing binary system within a globular
cluster would allow for a near direct distance determination to the
globular cluster (Paczy\'{n}ski 1997).  If the binary is well
detached and uncomplicated, accurate photometry and radial velocities
can be combined with a surface brightness-color relation to obtain the
distance to the globular cluster.  A number of authors have searched
for such binaries in globular clusters (e.g.\ Yan \& Mateo 1994,
McVean \ea\ 1997, Kaluzny \ea\ 1998).  McVean \ea\ (1997) have
identified one eclipsing binary system in the globular cluster M71
which appears to be a detached or semi-detached system, with the
detached model being more likely.  Detached eclipsing binary systems
have great potential as distance indicators to globular clusters which
will (hopefully) be realized in the next few years.

\section{Summary \label{sect7}}
The release of the Hipparcos data set has led a number of authors to
study the distance scale to globular clusters.  The Hipparcos data set
of high quality parallaxes for a number of nearby metal-poor stars has
renewed interest in the use of main sequence fitting to determine
distances to globular clusters.  In addition, the Hipparcos data on
proper motions of field RR Lyrae stars has been used to determine a 
new calibration of the absolute magnitude of the RR Lyrae stars
(via the statistical parallax method) which can be used to determine
the distances to globular clusters.  Over the last few years, a
variety of other methods have been used to derive  distances to 
globular clusters.  Given that many globular clusters contain RR Lyrae
stars, these distance determinations  can be compared 
via their calibration of the absolute magnitude of the RR Lyrae
stars.  This calibration is presented in equations (\ref{statpi}) --- 
(\ref{eqwhitefit}) and summarized in Table \ref{result}.
\begin{table}[htb]
\begin{center}
\begin{minipage}{6.56cm}
\caption{\mvrr\ at $\feh = -1.6$}
\label{result}
\begin{tabular}{ll}
\hline
\multicolumn{1}{c}{Method}&
\multicolumn{1}{c}{\mvrr}\\
\hline
Statistical Parallax & $0.77\pm 0.13$\\
Astrometric Distances& $0.60\pm 0.10$\\
White Dwarf Fitting  & $0.58\pm 0.10$\\
Theoretical HB models& $0.49\pm 0.10$\\
LMC                  & $0.46\pm 0.11$\\
Main Sequence Fitting& $0.45\pm 0.10$\\
\hline
\end{tabular}
\end{minipage}
\end{center}
\end{table}

The various calibrations  fall into three groups.  Main
sequence fitting using Hipparcos parallaxes, theoretical HB models and
the RR Lyrae in the LMC all favor a bright calibration, implying a
`long' globular cluster distance scale.  White dwarf fitting and the
astrometric distances yield a somewhat fainter RR Lyrae calibration,
while the statistical parallax solution yields faint RR Lyrae stars
implying a `short' distance scale to globular clusters.  The various
secondary distance indicators discussed in \S \ref{sect5} all favor
the long distance scale.  It is interesting to note that Hipparcos
provides support for both the long (from main sequence fitting) and
short distance scales (from statistical parallax).

A straight average of all six calibrations presented in Table
\ref{result} yields $\mvrr = 0.56$ mag with a standard deviation of 0.12 mag.
If the statistical parallax solution is removed from the average, then
$\mvrr = 0.52$ mag with a standard deviation of $0.07$ mag.  At the
present time, their is no reason to doubt the validity of the
statistical parallax solution.  A number of authors, using a variety
of data sources have all reached similar conclusions (see Layden, this
volume).  A possible explanation for the different result obtained
using statistical parallax compared to the other methods is that it is
the only method which calibrates the field RR Lyrae population (as
opposed to the RR Lyrae stars in a globular cluster).  Perhaps there
is a systematic difference between the field and globular cluster RR
Lyrae populations.  However, a study of the pulsation properties of RR
Lyrae variables in the field and in globular clusters found
essentially indistinguishable period-temperature distributions for the
two populations, suggesting that there is no significant difference in
luminosity between them (Catelan 1998).  For the above reasons, it
appears prudent at this time to include the statistical parallax
solution in the average.  This leads to a best estimate of the \mvrr\
calibration which can be used to set the globular cluster distance
scale of 
\begin{eqnarray}
\mvrr\ = (0.23\pm 0.04)(\feh + 1.6) +  (0.56\pm 0.12),
\label{eqfinal}
\end{eqnarray}

where the standard deviation among the six independent distance
techniques has been used as the error in the zero-point.
This is 0.1 mag fainter than that obtained from main sequence fitting,
but is 0.2 mag brighter than the statistical parallax solution.
Equation (\ref{eqfinal}) may be compared to my best estimate for the
the calibration of the RR Lyrae distance scale prior to the release of
the Hipparcos data  which implied $\mvrr = 0.66\pm 0.10$ mag at $\feh = -1.6$
(Chaboyer \ea\ 1996).  

The impact of this distance scale on the mean age of the oldest
globular clusters can be evaluated using the formulae presented by
Chaboyer \ea\ (1998) in the caption to their Figure 3.  From this,
equation (\ref{eqfinal}) implies a mean age of the oldest globular
clusters of $13\pm 2$ Gyr.  The dominant uncertainty in this age
estimate is the uncertainty in the distance scale to the globular
clusters.  In order to reduce the uncertainty in the absolute ages of
the globular clusters, the differences between the `long' distance
scale (based upon main sequence fitting, theoretical HB models and the
RR Lyrae in the LMC) and the `short' distance scale (based upon the
statistical parallax method) must be reconciled.


\vspace*{0.5cm}

\parindent 0pt

{\bf References}

\begin{description}
\item[]Brocato, E., Castellani, V.\ \& Piersimoni, A.\ 1997, ApJ, 491, 789
\item[]Buonanno, R., Corsi, C. E., Cacciari, C., Ferraro, F.R.\ \&
Fusi Pecci, F.\ 1994, A\&A, 290, 69
\item[]Buonanno, R., Corsi, C. E.\ \& Fusi Pecci, F.\ 1989, A\&A, 216, 80
\item[]Burnstein, D.\ \& Heiles, C.\ 1982, AJ, 87, 1165
\item[]Caloi, V., D'Antona, F.\ \& Mazzitelli, I.\ 1997, A\&A, 320, 823
\item[]Canuto V.M.\ \& Mazzitelli, I.\ 1991, ApJ, 370, 295
\item[]Caputo, F.\ 1997, MNRAS, 284, 994
\item[]Carney, B.W.\ 1979, AJ, 84, 515
\item[]Carney, B.W., Storm, J., Trammell, S.R.\ \&  Jones, R.V.\ 1992,
PASP, 104, 44
\item[]Carretta, E. \& Gratton, R.G.\ 1997, A\&AS, 121, 95
\item[]Cassisi, S., Castellani, V, Degl'Innocenti, S.\ \& Weiss, A.\
1998, A\&AS, 129, 267
\item[]Catelan, M.\ 1998, ApJ, 495, L81
\item[]Chaboyer, B.\ 1996, Nuclear Physics B Proceedings
Supplement, 51B, 10
\item[]Chaboyer, B., Demarque, P., Kernan, P.J.\ \& Krauss,
L.M.\ 1996, Science, 271, 957
\item[]Chaboyer, B., Demarque, P., Kernan, P.J.\ \& Krauss,
L.M.\ 1998, ApJ, 494, 96
\item[]Cudworth, K.M.\ 1979, AJ, 84, 1212
\item[]Fernley, J., Barnes, T.G., Skillen, I., Hawley, S.L., Hanley,
C.J., Evans, D.W., Solano, E.\ \& Garrido, R.\ 1998a, A\&A, 330, 515
\item[]Fernley, J., Carney, B.W.\ Skillen, I.\ Cacciari, C.\ \& Janes,
K.\ 1998b, MNRAS, 293, L61
\item[]Fusi Pecci, F., Buonanno, R., Cacciari, C.,
Corsi, C. E., Djorgovski, S. G., Federici, L., Ferraro, F. R., 
Parmeggiani, G., \& Rich, R. M.\ 1996, \aj, 112, 1461
\item[]Gould, A.\ \& Popowski, P.\ 1998, ApJ, in press
\item[]Gratton, R.G., Fusi Pecci, F., Carretta, E.,
Clementini, G., Corsi, C.E.\ \& Lattanzi, M.\ 1997, ApJ, 491, 749
\item[]Grundahl, F., VandenBerg, D.A.\ \& Andersen, M.I.\ 1998, ApJ,
500, L179
\item[]Harris, W.E.\ 1996, AJ, 112, 1487
\item[]Jimenez, R.\ \& Padoan, P.\ 1998, ApJ, 498, 704
\item[]Kaluzny, J, Hilditch, R.W., Clement, C.\ \& Rucinski, S.M.\
1998, MNRAS, 296, 347
\item[]Kov\'{a}cs, G., \& Jurcsik, J. 1996, \apj, 466, L17
\item[]Kov\'{a}cs, G., \& Walker, A.R.\ 1998, ApJ, submitted
\item[]Kraft, R.P., Sneden, C., Langer, G. E.\ \& Prosser, C.F.\ 1992, AJ,
104, 645
\item[]Kraft, R.P., Sneden, C., Smith, G.H., Shetrone, M.D., Langer, G.E.\ \&
Pilachowski, C.A.\ 1997, AJ, 113, 279
\item[]Lee, Y.-W. 1991 \apj, 373, L43
\item[]Lee, Y. -W., Demarque, P., \& Zinn, R. J. 1990, \apj, 350, 155 
\item[]Lee, Y. -W., Demarque, P., \& Zinn, R. J. 1994, \apj, 423, 248
\item[]McVean, J.R., Milone, E.F., Mateo, M.\ \& Yan, L.\ 1997, ApJ,
481, 782
\item[]Minniti, D., Geisler, D., Peterson, R.C.\ \& Claria, J.J.\ 1993, ApJ,
413, 548
\item[]Nissen, P., Gustafsson, B., Edvardsson, B.\ \&
Gilmore, G.\ 1994, A\&A, 285, 440
\item[]Paczy\'{n}ski, B. 1997, in The Extragalactic Distance
Scale, eds. M.\ Livio,
M.\ Donahue \& N.\ Panagia (Cambridge Univ. Press, Cambridge) 273
\item[]Pont, F., Mayor, M, Turon, C.\ \& VanDenberg, D.A.\
1998, A\&A, 329, 87
\item[]Rees, R.F.\ 1996, in Formation of the Galactic Halo .... Inside
and Out, eds.\ H.\ Morrison \& A.\ Sarajedini (San Fransico: ASP), 289
\item[]Reid, I.N., 1996, MNRAS, 278, 367
\item[]Reid, I.N., 1997,  AJ, 114, 161
\item[]Reid, I.N., 1998,  AJ, 115, 204
\item[]Renzini, A.\ 1991, in Observational Tests of Cosmological 
Inflation, eds.\ T.\ Shanks, \ea, (Dordrecht: Kluwer), 131
\item[]Renzini, A., Bragaglia, A., Ferraro, F.R., Gilmozzi, R., Ortolani, S.,
Holberg, J.B., Liebert, J., Wesemael, F.\ \& Bohlin, R.C.\ 1996, ApJ,
465, L23
\item[]Sandage, A. R. 1981a, \apj, 244, L23
\item[]Sandage, A. R. 1981b, \apj, 248, 161
\item[]Simon, N.R., \& Clement, C.M.\ 1993, ApJ, 410, 526
\item[]Sneden, C., Kraft, R.P., Prosser, C.F.\ \&  Langer, G. E.\ 1991, AJ,
102, 2001
\item[]Sneden, C., Kraft, R.P., Prosser, C.F.\ \&  Langer, G. E.\ 1992, AJ,
104, 2121
\item[]Walker, A.R.\ 1992, ApJ, 390, L81
\item[]Walker, A.R.\ 1994, AJ, 108, 555
\item[]Zinn, R.\ 1985, ApJ, 293, 424
\item[]Zinn, R.\ \& West, M.\ 1984, ApJS, 55, 45
\end{description}

\end{document}